# Code Definition Analysis for Call Graph Generation

Anne Veenendaal[1], Eddie Jones[1], Zhao Gang[1], Elliot Daly[1], Sumalini Vartak[2], Rahul Patwardhan[3]

[1]Computer Science and Emerging Research Lab, AU
[2]Lead IT, USA
[3]Infobahn Softworld Inc, USA

*Abstract*— Enterprise level software is implemented using multi-layer architecture. These layers are often implemented using decoupled solutions with millions of lines of code. Programmers often have to track and debug a function call from user interface layer to the data access layer while troubleshooting an issue. They have to inspect the code based on search results or use design documents to construct the call graph. This process is time consuming and laborious. The development environment tools are insufficient or confined to analyzing only the code in the loaded solution. This paper proposes a method to construct a call graph of the call across several layers of the code residing in different code bases to help programmers better understand the design and architecture of the software. The signatures of class, methods, and properties were evaluated and then matched against the code files. A graph of matching functions was created. The recursive search stopped when there were no matches or the data layer code was detected. The method resulted in 78.26% accuracy when compared with manual search.

## I. Introduction

Typical enterprise level application contains millions of lines of code. When a programmer inspects the code, he has to rely on search tools to analyze the relationship between various tiers of code. These results are based on the key words used by the programmer. The process of finding the call graph initiated in the user interface layer and tracking it all the way to the data access layer is a time consuming and laborious task. This paper examines the automation of call graph generation so that it is easier for programmers to understand how data flows across various layers of the architecture. A study [1] implemented and evaluated the performance of a program comprehension platform. It used code traversal technique to establish relation between various modules in the software. Another study [2] implemented software analysis while code rewriting process inside the JAVA language framework. Besides code traversal other strategies such as use of transforms to map code with various inter-linked models have been examined in studies [3] to make the software scalable.

Another study [4] used concepts from relational databases and applied them on an object-oriented language to develop relational programming. Researchers have developed utilities [5] and syntax for transforming programs to meaningful content. A study [6] used data about the project and codebase (meta-data) to create plugins for the eclipse integrated development environment. Another study [7] developed transformation rules and grammar to change the software code into human readable format. While most of these tools and techniques are focused on transformation of code, some studies [8], [9], [10], [11], [12] have used query-based approach for understanding the software architecture and code. There have been several studies [13] to [45] on text mining, emotion and sentiment mining from audio-visual data, text and other types of data with large size.

These studies have also discussed implementation schemes, design patterns and software architectural approaches to support large complex code bases and real time evaluation of continuous stream of large chunks of data. We believe the call graph generation strategies also have application in these other research areas in terms of text mining, which is similar to code mining except for the unstructured nature of the data. In this research we focus on c# based structured code on the back end and exclude the client side scripting syntax or any view specific syntax such as cshtml, xaml, html and javascript.

## II. Method

Four open source software products were chosen for the study. All the code bases were written in C#.NET and were web applications. 3 applications were implemented using web forms in ASP.NET and 1 application was implemented using MVC architecture. The automated call graph generation system consisted of 4 rules to match the class, method, property and layer transition. The first rule consisted of <class qualifier> <class name> <class inheritance>. The automation system searched for all the classes that matched the search string using the first rule.

The second rule consisted of : <accessibility qualifier> <return type> <method name> <method parameters>. The search engine looked for all matching functions with each class using second rule. The third rule was used for matching the properties in the class and was defined as: <access qualifier> <return type> <property name>. Once the search tree was generated, the code for each class, method, property was evaluated recursively to complete the call tree. Additionally the fourth rule to transition from one layer to the lower layer (user interface to business layer or business layer to data layer) was used to search into the next project code base. This rule was critical to provide a complete analysis of the call graph by crossing the project code-base boundary.

The automation process was evaluated on four solutions and compared with manual tracking of a functional call triggered from the user interface layer. The results of the accuracy are shown in the next section. To test the call graph accuracy, a function name was specified as the search criteria. The code analysis tool searched throught the code base folder for all matching files containing the search key word. It then categorized the files in code behind c# files and non-code files such as xml, xslt, html, resx etc. Each code file was displayed as the link. When the user clicked the link a complete call graph was generated to display the function calls originating from the functions top-down to the database layer.

The internal workflow of the code analysis tool and the algorithm is explained as follows: The code analysis tool consisted of two main components. The first component is responsible for maintaining an up to date linked dependency structure for all the objects, namespaces, classes, functions in the code. It is a back end job, which runs asynchronously at a schedules time, sweeps the code base, and regenerates the links. This was done to account for the constant code churn introduced by the maintenance and feature release development cycle. The second module was responsible for searching the tree by key word and finding the matching links in the tree and presenting the call graph to the user.

The first component responsible for structural mining of code consists of following modules discussed as follows:

1) Namespace extraction:

   The code analyzer detected the namespaces used in the class file. These namespaces are specified at the top of c# files and each line begins with keyword "using". The parser read each line and checked if it started with the word "using" and then tokenized the remaining portion of the string until it reached a semi-colon, which indicates an end of an expression in c#. The first token was always the "using" key word. The second token was the namespaces dependencies for the current class. Each of these tokens contained a dot (".") to indicate a category or folder. This was split further using the "." as the de-limiting character.

2) Class extraction:

   The class resolution module handled the next task of class resolution in the pipeline. This module relied on the knowledge that the classes preceded with the keyword "class". This keyword was used to tokenize the line containing the class declaration and the word following the keyword was extracted to derive the class name.

3) Inheritance extraction:

   The inheritance analyzer looked for the lines that had the class declaration and used the fact that the parent classes or interfaces were separated by the ":" (colon) as the keyword separator. The line was split into tokens containing the class declaration and the parent class and interface list separated by comma. This list was further split into individual classes and the interfaces, which were marked for traversal.

4) Composition extraction:

   The composition resolution code inspected the code files for data members of custom class data type and excluded the default language specific data types such as int, string, float etc. Once the lines were detected they were split into tokens using the space de-limiter and marked for traversal.

5) Function extraction:

   The function resolution component was the most important because it formed the links between various objects and entities in the code and function to function call connectivity. To detect a function call the analyzer looked for a dot operator "." followed by the token representing either the function name or a property name. If the token was followed by a "(" keyword then it was a function call vs the property getter or setter.

6) Web service call extraction:

   The web service call resolution was similar to function call resolution except for the fact that the code was not entirely available for navigation in the same project or class. The analyzer had to rely on the linking to navigate to the inner details of the web service code. In the current execution context, it only had access to the proxy code file.

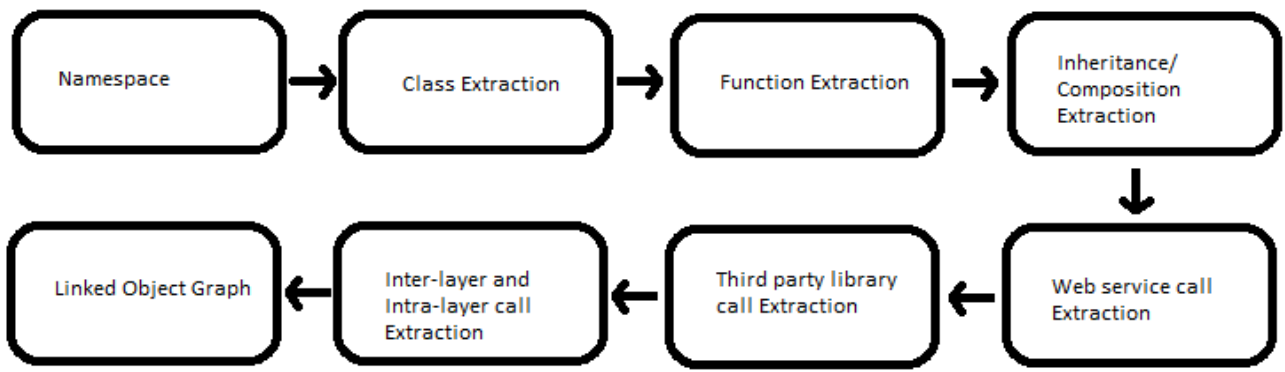

Fig. 1 Processing workflow for object and link extraction from structured object oriented code.

7) Third party assembly call extraction:

   The third party call resolution could only detect the call to third pary library but was limited due to lack of access to the code files. Since the strategy relied on source code analysis and not code disassembly, the third party call were the leaf nodes in the call graph tree.

8) Intra-layer call resolution:
   For the intra-layer call resolution the analyzer looked at the object.functioncall() expression and then checked the data type of the object making the call. If the object had the same namespace then it was treated as intra-layer call which meant call within the same layer such as business-business, UI to UI etc.

9) Inter-layer call resolution:
   For the inter-layer call resolution the analyzer looked at the object.functioncall() expression and then checked the data type of the object making the call. If the object had a different namespace and it belonged to a lower level layer call such as UI to business or business to data or UI, business or data to web service then it was treated as inter-layer call which meant call within the same layer such as business-business, UI to UI etc.

10) Static class resolution:
    The code to resolve the static class function calls used the same strategy of locating the object.functionCall() expression. The big difference here was that the static calls were made directly using the class type instead of using an object instance of a custom class type.

11) Anonymous functions:
    The anonymous functions were localized in nature and could be detected using the token for lambda expressions denoted by "=>". These functions formed the leaf nodes.

Once the first module completed the batch processing and object linking, the second module served as the call graph navigator when a search keyword was provided or a user interface level function was selected.

## III. RESULT

The call graph code structure analyzer component was executed daily as a scheduled job and the objects and functional links were evaluated for all four projects.

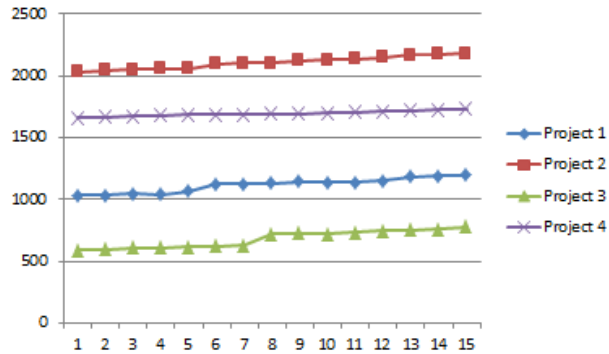

Fig 2. Daily code structural mining graph size by project.

The functional analyzer comparison for various projects per day is shown below. For all the projects the object linking identified steady increase in function count as the development code was checked into the code base.

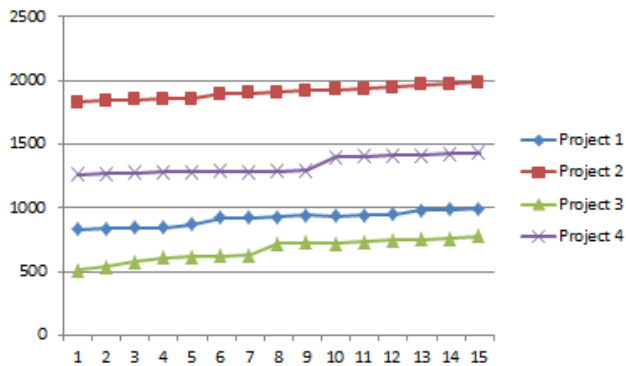

Fig. 3. Daily functional analyzer output by project.

At times there was sharp spike in the count because of high number of files getting added to the code base. But the code structure analyzer also detected dips in the function count, this was because of refactoring of code or removal of dead code or rollback of incorrect code.

Depending on the size of the project code base the functions detected in the function extractor module changed. But the discovered function count was consistent with the code base size and no anomalies were seen.

The call graph generation process was evaluated using the number of nodes (function calls) in the call graph chain. The count of functions detected using automated process was compared with the manually discovered function calls.

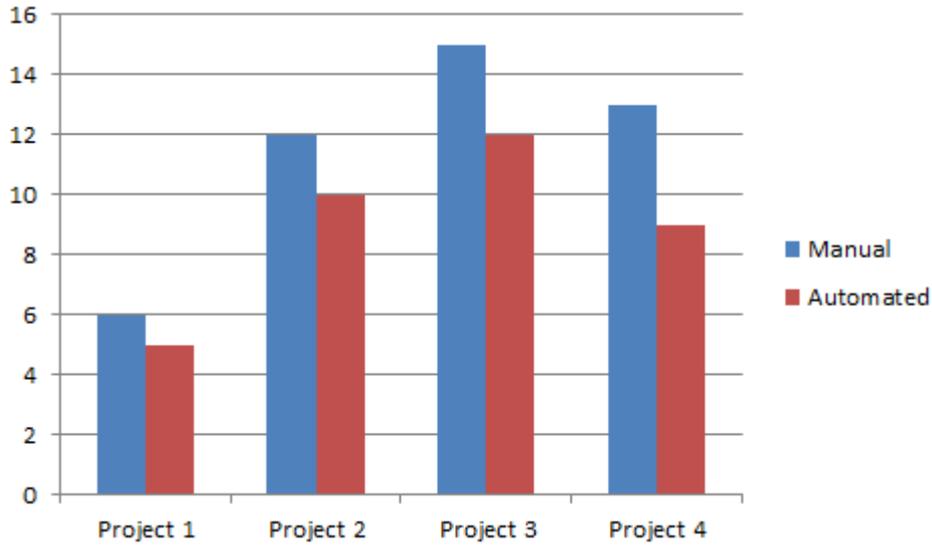

Fig. 4 Call graph node count manual vs automated.

The results indicated that the accuracy of the automated process in call graph node detection was 78.26%. The next metric used for evaluating the feasibility of the automation process was the time taken between manual discovery and the automated call graph generation.

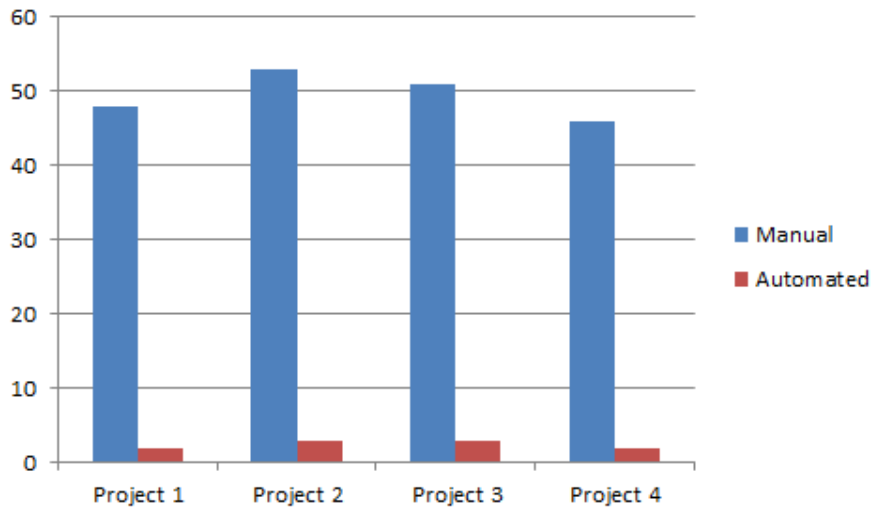

Fig. 5. Comparison of manual vs automated discovery time in minutes

The average discovery time taken by manual process was 49.5 minutes whereas the average time taken by automated call graph discovery process was 2.5 minutes. This showed that the automated process was definitely useful to improve productivity.

IV. CONCLUSIONS

The code definition analysis and signature inspection method resulted in 78.26% accuracy when compared with actual call graphs calculated by manual inspection. As a future scope analysis must be done on desktop applications and software products written in languages other than C#. Additionally, this study only focused on server side language and there is an opportunity to evaluate the call-graph generation method on client side scripting languages such as JavaScript and any other off shoots such as jquery and knockout. In terms of processing speed the automatic call graph discovery process was significantly faster (2.5 min vs 49.5minutes) than the manual process, thus indicating that focus should be given on improvement on the accuracy of the technique since the performance in terms of speed was already high enough to be acceptable for practical use. The modules in the call graph analyzer would have to be recoded to take into account different structure of client side code and support for method chaining which was not included as part of the server side code implementation. The same workflow engine also has potential applications in text mining frameworks and sentiment analysis.